\documentclass[twocolumn]{aastex631}
\usepackage{color,epsfig,amsmath}

\submitjournal{ApJ}

\shorttitle{Rings, shells, and arc structures around B[e] supergiants}
\shortauthors{D. H. Nickeler \& M. Kraus}

\begin{document}
\title{Rings, shells, and arc structures around B[e] supergiants:
I. Classical tools of non-linear hydrodynamics}

\correspondingauthor{Dieter H. Nickeler}
\email{dieter.nickeler@asu.cas.cz}

\author[0000-0001-5165-6331]{Dieter H. Nickeler}
\affiliation{Astronomical Institute, Czech Academy of Sciences, Fri\v{c}ova 298, 251\,65 Ond\v{r}ejov, Czech Republic}

\author[0000-0002-4502-6330]{Michaela Kraus}
\affiliation{Astronomical Institute, Czech Academy of Sciences, Fri\v{c}ova 298, 251\,65 Ond\v{r}ejov, Czech Republic}

\begin{abstract}
Structures in circumstellar matter reflect both fast processes and quasi-equilibrium states. A geometrical diversity of emitting circumstellar matter is observed around evolved massive stars, in particular around B[e] supergiants. We recapitulate classical analytical tools of linear and 
non-linear potential theory, such as Cole-Hopf transformations and Grad-Shafranov theory, and develop them further to explain occurrence of the circumstellar matter structures and their dynamics. We use potential theory to formulate the non-linear hydrodynamical equations and test dilatations of the quasi-equilibrium initial conditions. We find that a wide range of flow patterns can basically be generated and the time scales can switch, based on initial conditions, and lead to eruptive processes, reinforcing that the non-linear fluid environment includes both quasi-stationary structures and fast processes like finite-time singularities. Some constraints and imposed symmetries can lead to Keplerian orbits, while other constraints can deliver quasi-Keplerian ones. The threshold is given by a characteristic density at the stellar surface.
\end{abstract}

\keywords{hydrodynamics --- methods: analytical --- stars: winds, outflows --- circumstellar matter} 



\section{Introduction} \label{sec:intro}

During their post-main sequence evolution, massive stars ($M > 8\,M_{\sun}$) can undergo phases of enhanced 
mass loss and material ejections, which may lead to the formation of circumstellar shells and disks. One 
group of objects is particularly peculiar. These are the B[e] supergiants. The early findings by 
\citet{1985A&A...143..421Z} of a hybrid character of the UV and optical spectra have led to the
assumption of a two-component wind emanating from these objects with a classical line-driven wind in polar 
direction along with a much slower, cool and dense equatorial (disk-forming) outflow in equatorial 
direction. 

Intense infrared excess emission points to the presence of significant amounts of hot circumstellar 
dust \citep{1986A&A...163..119Z}, which imprints its specific emission features on the infrared 
spectra of these objects \citep{2010AJ....139.1993K}. The existence of a disk-shaped dusty structure 
has been inferred from optical linear polarization observations \citep{1992ApJ...398..286M, 
2001A&A...377..581M} and has later-on been reinforced for a few close-by Galactic objects by optical 
interferometric observations \citep{2007A&A...464...81D, 2011A&A...526A.107M, 2011A&A...528A..20B, 
2012A&A...548A..72C, 2012A&A...538A...6W}. 

A warm and dense circumstellar disk provides an ideal environment for the formation of molecules, and 
ro-vibrational emission from hot molecular gas has been detected in the near-infrared from CO 
\citep{1988ApJ...334..639M, 1996ApJ...470..597M, 2005A&A...436..653M, 2013A&A...549A..28K, 
2014ApJ...780L..10K, 2016A&A...593A.112K, 2020MNRAS.493.4308K,   
2012A&A...543A..77W, 2012MNRAS.426L..56O, 2013A&A...558A..17O, 2015AJ....149...13M, 
2015MNRAS.447.2459S, 2018MNRAS.480.3706K} as well as from SiO 
\citep{2015ApJ...800L..20K}, and in the optical possibly from TiO \citep{1989A&A...220..206Z, 
2016A&A...593A.112K, 2012MNRAS.427L..80T, 2018A&A...612A.113T} in a large number of objects 
\citep[for a detailed description we refer to the review by][]{2019Galax...7...83K}. The detection of 
significant enrichment of the disk material in the isotope $^{13}$C, traced by intense emission of the 
molecule $^{13}$CO, reinforces that the circumstellar matter of B[e] supergiants must have been 
released from the stellar surface and cannot be a remnant from star formation 
\citep{2009A&A...494..253K, 2010MNRAS.408L...6L}.

Several theoretical approaches have been presented to explain the formation of dense outflowing disks 
from B[e] supergiants. \citet{1993ApJ...409..429B} proposed that the winds emanating from the polar 
regions of rapidly rotating massive stars are bend towards the equator regions where they collide 
forming the so-called wind-compressed disk. This analysis considered a spherical shape of the star, 
but the inclusion of the non-radial forces occurring on the rotationally distorted stellar surface
seems to prevent the formation of a wind-compressed disk \citep{1996ApJ...472L.115O}.

Rapid stellar rotation facilitates another mechanism that might be considered, the rotation-induced 
bi-stability \citep{2000A&A...359..695P}. Due to the decrease in surface temperature from the pole to 
the equator caused by the rotation of the star (known as gravity darkening), the threshold temperature 
of $\sim 25\,000$\,K for recombination of Fe\,{\sc iv} into Fe\,{\sc iii} might be crossed. Because 
Fe\,{\sc iii} has significantly more lines suitable to drive the wind, a substantial increase in mass 
flux can be expected at lower temperature, that is, towards equatorial regions 
\citep{1999A&A...350..181V}. However, the density enhancement that can be achieved by this 
scenario remains a factor 10-100 below the expectations from observations. When the bi-stability 
mechanism is combined with the slow-wind solutions discovered by \citet{2004ApJ...614..929C}, 
the situation improves and a higher density contrast can be generated in the vicinity of the star 
\citep{2005A&A...437..929C}, but for the price of an equatorial wind velocity that is 10-20 times 
higher than what has been inferred from observations. 

The effects of gravity darkening have also been utilized in computations of the latitude-dependent 
ionization structure in the winds of B[e] supergiants, and \citet{2006A&A...456..151K} has shown that, 
despite the fact that rapid rotation alone leads to a lower wind density in equatorial directions as 
was shown by \citet{1996ApJ...472L.115O} and \citet{2001A&A...372L...9M}, the wind material 
(in particular hydrogen and elements with a similar ionization potential) can still recombine 
in the equatorial wind region of these luminous objects leading to a zone of neutral gas confined to 
the equatorial plane, in which also molecules and dust can form. The existence of such neutral (in 
hydrogen) zones (or disk-like structures) has been concluded from the analysis of the observed line 
luminosity of the [O\,{\sc i}] lines that arise in these disks \citep{2007A&A...463..627K}. Follow-up 
2D models revealed that such recombination scenarios require very high equatorial stellar mass-loss 
rates \citep{2008A&A...478..543Z}. In a different approach of 2D dense, viscous outflowing disks it 
was found that viscous heating dominates the innermost disk regions leading to extremely high 
temperatures within the disk mid-plane and to instabilities with significant waves or bumps
in density and temperature \citep{2018A&A...613A..75K}.
 
Many new observations have been carried out in the past years providing clearer insight with respect 
to the density distribution and the dynamics within the circumstellar (disk) matter. In particular, it 
has been found that the circumstellar material is confined in series of rings, arcs, or spiral-arm 
like structures revolving the central object on (quasi-)Ke\-plerian orbits, rather than 
being spread over a disk in the classical picture \citep[e.g.,][]{2010A&A...517A..30K, 
2016A&A...593A.112K, 2023Galax..11...76K, 2011A&A...526A.107M, 2012MNRAS.423..284A, 
2018MNRAS.480..320M, 2018A&A...612A.113T}. The arrangement of these rings is thereby unique for each 
object \citep{2018MNRAS.480..320M}, and each ring can have a different density that follows not 
necessarily the usual radial distribution expected in an outflow. 
Moreover, these rings can have gaps or inhomogeneities, and they can be either stable in time 
\citep{2016A&A...593A.112K, 2023Galax..11...76K} or display temporal variabilities 
\citep{2018MNRAS.480..320M, 
2018A&A...612A.113T} including fading \citep{2020MNRAS.493.4308K}, complete disappearance 
\citep{2014MNRAS.443..947L}, but occasionally also a sudden appearance of a new structure 
\citep{2012MNRAS.426L..56O} possibly caused by a pile-up of matter in a steadily decelerating outflow 
\citep{2010A&A...517A..30K}.
These findings indicate that the mass-loss from these stars is not a smooth process, but could be 
related to ejection phases, possibly triggered by instabilities acting in the strongly inflated 
envelopes of such massive and luminous objects \citep{1993MNRAS.263..375G, 
1993MNRAS.264...50K, 1996MNRAS.282.1470G}. 

Motivated by this great diversity of circumstellar environments of B[e] supergiants ranging from 
stationary density distributions in the form of rotating rings with sometimes alternating densities, 
or arc-like features, to decelerating equatorial outflows with sudden pileup of matter, and the 
deficiency of existing models to describe them, we develop in this paper new perspectives with 
different hydrodynamical scenarios that might help understanding their formation\footnote{Even 
if a complete formation scenario cannot be provided by this analysis, we present possible physical 
trigger mechanisms.}  and stationary structure. In particular, the formerly discussed scenarios and 
phenomena should be interconnected with basic, generic properties of fluid dynamics. We want to 
consider the following open questions from an abstract perspective: 

\begin{itemize}
\item How can a stationary, ideal fluid representation (without dissipative
effects) of persistent matter (stable, quasi-stationary rings, arcs, complete or incomplete spiral 
structures) be constructed?
 
\item How can quasi-stationary mass distributions and probably time-dependent flows (e.g. in the form 
of episodic mass loss of the central star) appear together?

\item How can simplified single-fluid time-dependent and time-independent (stationary) velocity fields 
be constructed for such above mentioned cases, if not many detailed physical parameters are known?

\item Do such time-dependent velocity fields exist {\it at all for a stationary density distribution}?
What is their nature?
\end{itemize}

The paper is structured in the following way: In Section~\ref{sect:ideal} we present the stationary 2D 
solution techniques but also the transition to non-stationary and 3D problems within incompressible 
ideal hydrodynamics (HD). In Section~\ref{sect:potential} we derive a non-linear Schr\"{o}dinger-type 
equation as equivalent formulation of the compressible HD equations. In Section~\ref{sect:flow} we 
analyze a general 3D compressible flow on the basis of stationary stellar wind solutions. Our results 
are discussed and our conclusions are summarized in Section~\ref{sect:concl}.

\section{Incompressible hydrodynamics and blow-up solutions}\label{sect:ideal}

\subsection{Basic equations of ideal hydrodynamics}

The scenarios described in the introduction will be treated by using the 
basic equations of HD. These are given by the mass continuity equation (Equation\,(\ref{mce})) and the 
Euler equation (Equation\,(\ref{ee})),
\begin{eqnarray}
  \frac{\partial\rho}{\partial t} + {\boldsymbol\nabla\boldsymbol\cdot}\left( \rho\textbf{\textit{v}} \right) &=&
0\, ,\label{mce}\\
   \rho\left(\frac{\partial\textbf{\textit{v}}}{\partial t}+\left(\textbf{\textit{v}}{\boldsymbol\cdot\boldsymbol\nabla}\right)
\textbf{\textit{v}} \right) &=& -{\boldsymbol\nabla} p + \rho\textbf{\textit{g}}\, \label{ee},
\end{eqnarray}
in which $\rho$ is the mass density, $\textbf{\textit{v}}$ the gas velocity, $p$ is the gas pressure 
which may include also the radiation pressure given that it can be assumed to be isotropic\footnote{The radiation pressure can be assumed to be isotropic in the circumstellar matter around B[e] supergiants, which consists of significantly optically thick material. Otherwise, 
the non-isotropic part of the radiation pressure is neglected in our pure HD model.}, and 
$\textbf{\textit{g}} = -{\boldsymbol\nabla}\phi$ is the gravitational acceleration of the star with 
the gravitational potential $\phi$. Self-gravitation effects of the circumstellar 
matter are neglected. We focus on systems in which viscosity can be neglected, 
meaning that the length scales of the pressure force are much larger than the so-called deflection 
length, and the flows are super-sonic but not highly super-sonic \citep[i.e. no shocks 
involved, see][]{1992apa..book.....F}.

To allow for a profound investigation of the physical aspects, we will split our analysis into two 
distinct physical extremes: incompressible velocity fields (Section~\ref{incomp}) and irrotational 
potential velocity fields (Section~\ref{sect:potential}). One should be aware of the fact that
the incompressibility condition ${\boldsymbol\nabla\boldsymbol\cdot}\textbf{\textit{v}}=0$ is neither 
valid for classical viscous disks models, nor for classical wind solutions, as these usually 
assume isothermy. Our incompressibility model is not isothermal, and it makes shocks less likely to 
occur.

	\subsection{Stationary 3D incompressible flows}\label{incomp}

An interesting case that has emerged from the observations are the (presumably) Keplerian rotating 
rings detected around the B[e]SG star LHA 120-S 73 \citep{2016A&A...593A.112K}. These rings did not 
show any measurable radial motion within the 
observation period spanning 16 years, justifying the assumption of a quasi-stationary model, i.e., 
time-dependent changes of physical quantities are small and can be neglected. Moreover, due to the 
preferential rotational and relaxed motion of the gas along closed orbits, the compressibility of the 
gas is negligible meaning that ${\boldsymbol\nabla\boldsymbol\cdot}\textbf{\textit{v}}=0$. 
In this case, the mass conservation Equation\,(\ref{mce}) simplifies to 
\begin{equation}
\textbf{\textit{v}}{\boldsymbol\cdot\boldsymbol\nabla}\rho=0
\end{equation}
implying that the flow is always perpendicular to the gradient of the density 
($\textbf{\textit{v}}\perp{\boldsymbol\nabla}\rho$), respectively the density is constant along 
streamlines. 

We now introduce the streaming vector \citep[or auxiliary flow vector,][]{2005AdSpR..35.2067N, 2006A&A...454..797N} $\textbf{\textit{w}}$ via
\begin{equation}
\textbf{\textit{w}} = \sqrt{\rho}\textbf{\textit{v}} \label{stream}
\end{equation}
so that 
\begin{equation}
 \textbf{\textit{w}}{\boldsymbol\cdot\boldsymbol\nabla}\rho=0\quad\land\quad
 {\boldsymbol\nabla\boldsymbol\cdot}\textbf{\textit{w}}=0\, . \label{imc}
\end{equation}
%
The total pressure, or as we will call this in the following, the Bernoulli-pressure $\Pi$, 
is defined by
\begin{equation}
\Pi=p+\frac{1}{2}\rho \textbf{\textit{v}}^2  = p  + \frac{1}{2} \textbf{\textit{w}}^2\,. \label{ber}
\end{equation}
%
Applying the identity
\begin{equation}
\left(\textbf{\textit{w}}{\boldsymbol\cdot\boldsymbol\nabla}\right) \textbf{\textit{w}} = 
\boldsymbol\nabla \left(\frac{1}{2} \textbf{\textit{w}}^{2} \right) - \textbf{\textit{w}}{\boldsymbol\times}\left({\boldsymbol\nabla}{\boldsymbol\times}\textbf{\textit{w}}\right) \label{Weber}
\end{equation}
to the stationary Euler Equation\,(\ref{ee}) and the 
definitions for the streaming vector in Equations\,(\ref{stream})-(\ref{ber}),  the resulting momentum equation in 3D can be written in the form
\begin{equation}
{\boldsymbol\nabla}\Pi = \textbf{\textit{w}}{\boldsymbol\times}\left({\boldsymbol\nabla}{\boldsymbol\times}\textbf{\textit{w}}\right) - \rho{\boldsymbol\nabla}\phi\, . \label{iee}
\end{equation}
The Equations (\ref{imc}) and (\ref{iee}) form the set of incompressible HD equations that need to be 
solved. 

\subsubsection{Solutions in 2D}\label{sol2d}

Due to flat or disk-like structuring and concentration of the matter around the star, we assume
that the material is confined in a thin sheet around the equatorial plane. In 
general, circumstellar disks can be assumed to be symmetric around the mid-plane, having their 
maximum density and pressure along $z=0$ so that $\partial/\partial z = 0$. Our analysis is restricted 
to the $xy$-plane and fringe effects are neglected. The investigated scenario is not intended to 
generate an outflowing disk in the classical sense, but to find representations for revolving rings 
or arcs in a quasi-stationary state. In the limit, we assume a purely azimuthal flow and use a 
different geometry (basically 2D in Cartesian components, i.e. $(x,y)=(R,\Phi)$), where in contrast 
most works about disks use the $(R,z)$ coordinate system (i.e. $\partial/\partial \Phi = 0$). 

To satisfy the mass conservation Equation (\ref{imc}) we define the stream function $\psi=\psi(x,y)$, 
the streaming vector $\textbf{\textit{w}} = {\boldsymbol\nabla}\psi {\boldsymbol\times} 
\textbf{\textit{e}}_{z}$, and the mass density $\rho = \rho(\psi)$. 
In the general case, ${\boldsymbol\nabla}\psi$ and
${\boldsymbol\nabla}\phi$ are not parallel to each other almost everywhere, such that
the stream function $\psi$ and the gravitational potential $\phi$
can be regarded as coordinates replacing $x$ and $y$,
and $\Pi$ is consequently an explicit function of $\psi$ and $\phi$.
Inserting these relations into the momentum 
Equation (\ref{iee}) and expanding with respect to the basis vectors ${\boldsymbol\nabla}\psi$ and 
${\boldsymbol\nabla}\phi$ delivers
\begin{eqnarray}
{\boldsymbol\nabla}\Pi &&=\Delta\psi{\boldsymbol\nabla}\psi-\rho(\psi){\boldsymbol\nabla}\phi\, ,\\
\Rightarrow \frac{\partial\Pi}{\partial\psi}{\boldsymbol\nabla}\psi+\frac{\partial\Pi}{\partial\phi}
{\boldsymbol\nabla}\phi
&&=\Delta\psi{\boldsymbol\nabla}\psi -\rho(\psi){\boldsymbol\nabla}\phi\, .
\end{eqnarray}
From a comparison of coefficients we obtain one non-linear Poisson-like  partial differential 
equation for the stream function and one equation for the external gravitational potential
\begin{equation}
\frac{\partial\Pi}{\partial\psi}=\Delta\psi\quad\land\quad
\frac{\partial\Pi}{\partial\phi}=-\rho(\psi)\, . \label{pdes}
\end{equation}
This set of relations is an analogy to results in magnetohydrostatic (MHS) theory  with
gravity \citep{1983SoPh...87..103S}, while the first quasi- or non-linear elliptic type equation has
already been found and described in HD by \citet{1848TCaPS...7..439S}.
Formal integration of the right-hand equation of Equation (\ref{pdes}) results in a relation for the 
Bernoulli-pressure
\begin{equation}
\Pi = -\rho(\psi)\, \phi + \Pi_{0}(\psi)\, , \label{pi_int}
\end{equation}
however, this formal integration does not solve the system completely.
Inserting this Bernoulli-pressure into the left-hand equation of Equation (\ref{pdes}) delivers
\begin{equation} 
\Delta\psi=-\rho'(\psi)\phi+\Pi_{0}'(\psi)\, ,\label{lse1}
\end{equation}
where the prime denotes the derivative with respect to $\psi$.

To be able to solve this equation, it is necessary to know the function 
$\rho(\psi)$, i.e. the density as a function of the stream function.
While the stream function $\psi$ is not known a priori, the density along
each stream-line label (i.e. the value of the stream function) must be constant,
but can vary across stream lines. With this knowledge, we are able to
calculate the 2D density structure, as we can basically choose the density function 
$\rho=\rho(\psi)$ arbitrarily. The choice of the density function implies
a non-linear feedback on the solution of the non-linear Laplace 
Equation\,(\ref{lse1}). The solution of this Laplace equation finally delivers
a stream function $\psi = \psi(x,y)$, based on which the spatial density 
distribution $\rho=\rho(x,y)$ can be computed\footnote{We will show in Section~\ref{Sect:proof} 
that this notion and the procedure can be transferred also to the limiting case, for which 
${\boldsymbol\nabla}\psi{\boldsymbol\times}{\boldsymbol\nabla}\phi={\boldsymbol 0}$, i.e. for one-dimensional equilibria.}.

Taking the most simple, nontrivial approach given by
\begin{equation}
\rho'(\psi)=-\lambda = {\rm const}\, , \label{monotc}
\end{equation}
and assuming that $\Pi_{0}$ is constant, results in
\begin{equation}
  \frac{\partial\Pi}{\partial\psi}=\Delta\psi=\lambda\phi\, . \label{Poisson}
\end{equation}
The relation between the density and the stream function, Equation\,(\ref{monotc}), is chosen in such 
a way that the monotonicity of the density function, and therefore a unique relation, i.e. the 
bijective character of the density function, is guaranteed. We will recognize in the following that 
$\lambda$ controls the influence of the gravitation on the geometry and dynamics of the flow: the
larger $\lambda$, the larger the deviation from ordinary potential flows (see Equations\,(\ref{gs1}) 
and (\ref{wvec}) below).

To derive this influence we first of all have
to facilitate the solution of Equation (\ref{Poisson}), by switching to Wirtinger calculus \citep[e.g.][]{Remmert} and using the coordinates (or variables) $u:=x+iy$ and $v:=x-iy$ with $i^2=-1$. Then the Poisson equation,
Equation\,(\ref{Poisson}) can be written as
\begin{equation} 
4\partial_u\partial_v\psi=-\lambda\,\frac{GM_{*}}{\sqrt{uv}}\, ,
\label{Poisson2}
\end{equation}
where we inserted the definition of the gravitational potential given by $\phi = \frac{-GM_{*}}{R}$, 
in which $G$ is the gravitational constant, $M_{*}$ is the mass of the central star, and where $R$ defined 
as $R^{2} = uv$ is the radial distance from the center of the star within the $x$--$y$-plane,
with $R \ge R_{*}$, where $R_{*}$ denotes the radius of the star.

Integration of the solution of the Poisson Equation\,(\ref{Poisson2}) with respect to the $u$--$v$ coordinates delivers the stream function
\begin{equation}
\psi = -\left(\lambda GM_{*}\right)\sqrt{uv}+\psi_{1}(u)+\psi_{2}(v)\, ,
\label{gs1}
\end{equation}
where $\psi_{1}$ and $\psi_{2}$ are free functions.
With this stream function we can compute the streaming vector and the Bernoulli-pressure, where the thermal pressure $p$ can then be calculated by subtracting the kinetic pressure.
The general solution (Equation\,(\ref{gs1})) shows an inhomogeneous part depending on the coupling constant $\lambda$.

This simplified approach leads in a mathematical limit to a quasi-Kepler rotation. We assume that the 
homogeneous part of the general solution (Equation\,(\ref{gs1})), i.e. the meromorphic part, is zero.
Therefore the solution for $\psi$ and the derived
streaming vector can be used to calculate the (rotational) velocity of the gas around the star
\begin{eqnarray}
\textbf{\textit{w}} &=& -\left(\lambda GM_{*}\right){\boldsymbol\nabla} R
{\boldsymbol\times}\textbf{\textit{e}}_z \nonumber\\
&=& -\left(\lambda GM_{*}\right)\left[-\frac{x}{R}\textbf{\textit{e}}_y + 
\frac{y}{R}\textbf{\textit{e}}_x\right] \label{wvec}\\
\Rightarrow\,\,\, \textbf{\textit{v}} &=&  \frac{\textbf{\textit{w}}}{\sqrt{\rho(\psi)}}
=\frac{-\lambda GM_{*}}{\sqrt{\lambda^2 GM_{*} R+\rho_0}}\,\left(\frac{y}{R}, -\frac{x}{R}\right)
\\
\nonumber\\
\Rightarrow\,\, v^2 &=& \displaystyle\frac{GM_{*}}{R + \frac{\rho_0}{\lambda^2 GM_{*}}}=
\frac{GM_{*}}{R}\,\frac{1}{1+\frac{\rho_0}{\lambda^2 GM_{*} R}}
\end{eqnarray}
where we have inserted for the density function the solution of Equation\,(\ref{monotc}), i.e., 
$\rho(\psi)=-\lambda\psi+\rho_0=\lambda^2 G M_{*}\, R + \rho_0$. For the limit
$\lambda,\rho_0\rightarrow 0$ and $\rho_0/\lambda^2\rightarrow 0$, or for $\rho_0=0$,
the velocity is identical to the Kepler rotation. For a vanishing integration constant $\rho_0$, 
the density at the stellar surface is given by $\rho(R_*) = \lambda^{2} G M_{*} R_{*}$. While the 
density of the circumstellar matter in the vicinity of the star can take small values, depending on the choice of
$\lambda$, it is not trivial to generate a Keplerian rotating disk or ring detached from the 
stellar surface. 

Other cases, 
e.g., for large values of $\rho_0/\lambda^2$, lead to strong deviation from Kepler rotation close to 
the star. But, even in the latter case, the velocity can approach Keplerian behavior for sufficiently 
large values of $R$.  It should be noted that our approach only represents the close-by 
circumstellar environment, i.e., regions between the stellar surface and the first ring or onset of 
the disc. The linear function chosen for the density $\rho(\psi)$ might be regarded as the lowest 
order part of a Taylor expansion in the region of our interest of a complex density 
function, and is only meant to have a prescription from a low density, e.g. close to the star (\lq gap\rq), to high values (ring/disk). The lack of information of the real radial density distribution 
from observations hampers a more precise theoretical description. 

Inserting the streaming vector, Equation (\ref{wvec}), and the density function $\rho(\psi)$ into the 
equation for the pressure function, Equation (\ref{pi_int}), Equation (\ref{ber}) delivers
\begin{equation}
p = \frac{G M_{*}}{R} \rho_{0} + \frac{1}{2} \lambda^{2} G^{2} M_{*}^{2} + \Pi_{0}\, .
\end{equation}
If $\rho_{0}$ would be zero, the density would be given by $\lambda^2 G M_{*}\, R$, the pressure 
gradient (and therefore the pressure force) would vanish, and the motion of the gas would be purely
ballistic (Keplerian).

\subsubsection{On the existence of $\Pi(\psi,\phi)$ for one-dimensional equilibria}
\label{Sect:proof}

In Section \ref{sol2d} we showed for two-dimensional equilibria that the pressure $\Pi$ can be expressed as a function of the scalar fields $\psi$ and $\phi$ with
\begin{eqnarray}
\frac{\partial\Pi}{\partial\psi} &=& \Delta\psi:=-\Omega_w(\psi,\phi)\label{vort}\\
\nonumber\\
\frac{\partial\Pi}{\partial\phi} &=& \Delta\phi:=-\rho(\psi)\label{pois}\, ,
\end{eqnarray}
where $\Omega_w$ is the vortex strength (or vorticity in 2D) associated with the flow vector $\textbf{\textit{w}}$, which is defined by $\boldsymbol\Omega_{w} := {\boldsymbol\nabla}{\boldsymbol\times}{\boldsymbol w}$ and reduces in our case to $\boldsymbol\Omega_{w} = \Omega_{w} \textbf{\textit{e}}_z$. 

In the one-dimensional case, e.g. $\partial/\partial\Phi=0$, where $(x,y)\equiv(R,\Phi)$ (with $R$ and $\Phi$ being the radial coordinate and the azimuthal angle, respectively),
it is not obvious at first that there is an incompressible stationary solution, characterized 
by $\Pi(R)$, $\psi(R)$ and $\phi(R)$, the vorticity $\Omega_w(R)$, and the mass density 
$\rho(R)$.

A possible proof has been proposed by Hornig (1996, private communication) and presented by 
\citet{Fleischer}, which we recapitulate for clarification:
We consider the two-dimensional space $\rm{I\! R}^2$ with the coordinates $(\psi,\phi)$, in which the solution $\psi(R),\phi(R)$ is a curve $L$, parameterized by $R$. On $L$ the vector field $\textbf{\textit{V}}=V_\psi(\psi,\phi)\,\textbf{\textit{e}}_{\psi}+V_\phi(\psi,\phi)\,\textbf{\textit{e}}_{\phi}$ is given by
\begin{equation}
V_{\psi}(\psi(R),\phi(R))=-\Omega_w
\end{equation}
and
\begin{equation}
V_{\phi}(\psi(R),\phi(R))=-\rho \, .
\end{equation}
We are looking for a potential $\Pi$ on $\rm{I\! R}^2$ such that $\textbf{\textit{V}}=
{\boldsymbol\nabla}\Pi|_L$.

We assume that there is a $\varepsilon_0$-hose around $L$ that does not overlap anywhere.
Then there are local coordinates $(s,\varepsilon)$, where $s$ is the arc length along $L$ and $\varepsilon$ is the distance from $L$. These are orthogonal to $L$, $\textbf{\textit{e}}_s{\boldsymbol\cdot}\textbf{\textit{e}}_{\varepsilon}|_L=0$. There is then a decomposition of $\textbf{\textit{V}} (s)= V_s (s) \textbf{\textit{e}}_s + V_{\varepsilon} (s) \textbf{\textit{e}}_{\varepsilon }$ on $L$, and the potential $\Pi$, with $\Pi\equiv 0$ outside the tube, and
\begin{equation}
\Pi(\varepsilon,s)=\left[\int\limits_{0}^{s} \, V_s(\tilde{s})\,
d\tilde{s} + \varepsilon V_{\varepsilon}(s)\right]\exp\left(-\,\frac{
\varepsilon^2}{\varepsilon_{0}^2-\varepsilon^2}\right)
\end{equation}
for $\varepsilon\leq\varepsilon_0$
fulfills the requirement $\textbf{\textit{V}}={\boldsymbol\nabla}\Pi|_L$ inside the tube.

Thus there is a pressure function $\Pi(\psi,\phi)$ whose partial derivatives
according to the Equations\,(\ref{vort}) and (\ref{pois}) agree with the vorticity $\Omega_w$ and the mass density $\rho$ as functions of $\psi$ and $\phi$.

\subsubsection{Preliminary discussion: Do time-dependent incompressible flows exist in 2D or in 3D ?}

Even if a great variety of geometrical structures with regard to flow patterns can be generated 
in 2D with the method described in Section~\ref{sol2d}, we know that even quasi-stationary structures 
have to undergo a formation phase that can be quite eruptive and therefore strongly time-dependent. 
This leads to the question whether quasi-stationary structures can be preserved in case of a 
non-linear time-dependent change of the system or relax respectively blow up. Or, equivalently, is it 
at least possible to extend the stationary models via \lq slight amplitude modulation\rq, i.e. 
the assumption that stationary fields can be multiplied with a specific purely time-dependent 
function, to deliver a time-dependent solution including the stationary solution? To answer 
this question we focus on the dynamical behavior in the non-linear case. We 
additionally assume that the system reacts instantaneously on non-local perturbations.

The slight amplitude modulation is equivalent to posing the question if a time-separability from 
the spatial components guarantees regular solutions within a small, finite time interval. The
regularity would at least make it plausible that the solutions are non-linearly stable. The  
separability of solutions of evolutionary equations with respect to time is possible for many examples 
in continuum and quantum physics \citep[see, e.g.,][]{galaktionov2006exact}. The general solutions of 
such kind of equations can then be constructed by linear superposition of multiple summands 
constructed by such a separation ansatz, or by non-linear superposition of a similar
set of summands.

We assume a time-separability and take the general incompressible
3D representation into account. Considering the Euler equation for vanishing time derivative 
$\partial\rho/\partial t\approx 0$ and external force, and using the definition of the stream 
vector (Equations\,(\ref{stream}) and (\ref{imc})), we obtain
\begin{eqnarray}
{\boldsymbol\nabla} p &=& -\rho\left(\textbf{\textit{v}}{\boldsymbol\cdot\boldsymbol\nabla}\right)\textbf{\textit{v}}
-\rho\frac{\partial\textbf{\textit{v}}}{\partial t}\nonumber\\
\Leftrightarrow\quad {\boldsymbol\nabla}\Pi &=&\textbf{\textit{w}}{\boldsymbol\times}\left({\boldsymbol\nabla}{\boldsymbol\times}\textbf{\textit{w}}\right)
-\sqrt{\rho}\,\frac{\partial\textbf{\textit{w}}}{\partial t} \, .
\label{eu3}
\end{eqnarray}
With $\textbf{\textit{w}} = w_{1}(t)\, \textbf{\textit{w}}_{0}(\textbf{\textit{x}})$, and analogously for the total pressure
$\Pi$, we can rewrite the second form of the Euler Equation~(\ref{eu3}) as
\begin{equation}
\Pi_{1}(t)\,{\boldsymbol\nabla}\Pi_0({\textbf{\textit{x}}})=  w_{1}^2\,\textbf{\textit{w}}_0{\boldsymbol\times}\left({\boldsymbol\nabla}{\boldsymbol\times}
\textbf{\textit{w}}_0\right) - \dot{w}_1\,\sqrt{\rho}\,\textbf{\textit{w}}_0  \, ,
\end{equation}
where the dot in $\dot{w}_1$ denotes the derivative with respect to time. 
Assuming that $\exists\, (\textbf{\textit{w}}_0,\Pi_0, \rho)$ with
\begin{equation}
 {\boldsymbol\nabla}\Pi_0 =
 \textbf{\textit{w}}_0{\boldsymbol\times}\left({\boldsymbol\nabla}{\boldsymbol\times}\textbf{\textit{w}}_0\right)
-\sqrt{\rho}\,\textbf{\textit{w}}_0\label{eu4}
\end{equation}
delivers
\begin{equation}
\Pi_{1} = w_1^2 \propto \dot{w}_1 \quad\Rightarrow\quad
 w_1=\frac{w_{10}}{t_{0}-t}\,\, ,
\end{equation}
with the integration constant $w_{10}$. For $w_{10} > 0, t_0 > 0$ and $t_0 > t$ this solution 
constitutes a blow-up solution or finite-time singularity for $t\rightarrow t_0$, whereas in the 
limit $t\rightarrow -\infty$ the velocity and pressure decay. 
We would like to emphasize that Equation\,(\ref{eu4}) does {\it not} represent an equilibrium 
solution, but poses an additional constraint for the stationary part of the flow, respectively the 
steady-state flow pattern.

If we take also gravity into account for the above calculations then the time-independent form of the momentum equation is
\begin{eqnarray}
{\boldsymbol\nabla}\Pi_0 &&=
 \textbf{\textit{w}}_0{\boldsymbol\times}\left({\boldsymbol\nabla}{\boldsymbol\times}\textbf{\textit{w}}_0\right)
-\sqrt{\rho}\,\textbf{\textit{w}}_0- \frac{GM_{*}\rho}{r^2} \,\textbf{\textit{e}}_{r}\nonumber\\
\Rightarrow \quad \Pi_{1} &&= w_1^2=\dot{w}_1 \propto M_{*}(t)
\label{eu5}
\end{eqnarray}
where $\textbf{\textit{e}}_{r}$ is the unit vector directing radially outwards from the center 
of gravity, and 
$M_{*}(t)$ is the stellar mass changing over time. For $t_{0} < 0$ and $t=0$ the relation
$\dot{w}_1 = w_{10}/t_{0}^{2} = 1$ fulfills the condition that $M_*(t=0) = M_*$. Then it follows
that for $t>0$ and $t\rightarrow \infty$ the mass, pressure and velocity decrease, whereas they 
increase for $t_{0} > 0, t_{0} > t$ and $t\rightarrow t_0$. While the former can be interpreted 
with mass loss, the latter would imply mass accretion. It should be noted that both scenarios only 
hold for a meaningful (small) time interval, as the temporal variation of the velocity field is given 
by a (pure) dilatation.

The assumptions leading to Equations\,(\ref{eu3})--(\ref{eu5}) reflect a nonlinear perturbation of a 
system, whose original stationary (geometrical) structure should be preserved, whereas the 
amplitudes of the stationary field components should vary only slowly. The calculations show that the 
time-independent field components can in fact no longer satisfy any stationary equation, and that even 
the time-dependent amplitudes diverge, i.e. develop finite-time singularities, or decay
completely. Although the dynamics is 
limited to incompressible flows in these considerations, the behavior of the time-dependent amplitudes 
implies an eruptive mass loss (or mass gain) of the star. This means that within a short period of time the 
stellar mass loss (or gain) would be considerably enhanced when the system is forced by non-linear 
variations (perturbations) which can change the character of the system, driving it away from a
relaxed incompressible flow. Our analysis shows that time-dependent incompressible
flows around a stationary state cannot exist but they lead either to a completely decaying flow or 
to a blow up.

\section{Potential flows in 3D}\label{sect:potential}

In contrast to the studies in Section \ref{incomp}, the time-dependent change in velocity
field is now included. Also, we drop the restriction to incompressible
flows. Instead, we assume that the flow is irrotational. Moreover, we replace the
stream function with a scalar potential, which allows us to easily implement the time-dependence of the velocity field. 

In contrast to the stream function model used in Section \ref{incomp}, reflecting the isocontours
of the density distribution in a relaxed flow, we now utilize a scalar velocity potential. The
surfaces of constant potential represent a family of surfaces, which has an affinity to the family of surfaces of constant radial coordinate. This would favor a radial outflow, instead of the more 
azimuthal flows described by the stream function model in the previous section. In case of a time dependency, the potential describes expanding (or eventually shrinking) surfaces, to which the velocity vector is perpendicular, thus pointing in a certain sense \lq radially\rq~outwards (or inwards). The density, decreasing outwards (eventually inwards), should be therefore, at least locally, diffeomorphic with respect to the potential, as will be introduced and explained in the following.

First of all, maintaining the condition of a stationary density distribution, the equation of mass conservation reduces again to 
\begin{equation}
{\boldsymbol\nabla\boldsymbol\cdot}\left( \rho\textbf{\textit{v}} \right) = 0\, . \label{mpf}
\end{equation}
The Euler Equation\,(\ref{ee}) can be re-written, using the identity Equation\,(\ref{Weber}), as 
\begin{equation}
\frac{\partial\textbf{\textit{v}}}{\partial t}+{\boldsymbol\nabla}\frac{\textbf{\textit{v}}^2}{2} - \textbf{\textit{v}}{\boldsymbol\times}
\left({\boldsymbol\nabla}{\boldsymbol\times}\textbf{\textit{v}}\right)=-\frac{{\boldsymbol\nabla} p}{\rho}-{\boldsymbol\nabla}\phi\, . \label{epf}
\end{equation}
Introducing a 3D velocity potential $\textbf{\textit{v}}:={\boldsymbol\nabla}\varphi$, leading to ${\boldsymbol\nabla}{\boldsymbol\times}\textbf{\textit{v}} 
= 0$, and utilizing a barotropic law $p=p(\rho)$, Equations (\ref{mpf}) and (\ref{epf}) take the 
following form
\begin{eqnarray}
& & {\boldsymbol\nabla}\varphi{\boldsymbol\cdot}{\boldsymbol\nabla}\rho  = -\rho\Delta\varphi\, , \label{mbaro} \\
& & {\boldsymbol\nabla}\left(\int\frac{dp}{\rho(p)} +\frac{({\boldsymbol\nabla}\varphi)^2}{2} +
 \frac{\partial\varphi}{\partial t}+\phi\right)  =  {\boldsymbol 0}\, \nonumber \\ 
&\Leftrightarrow & \qquad \,\,\, \int\frac{dp}{\rho(p)} +\frac{({\boldsymbol\nabla}\varphi)^2}{2} +
 \frac{\partial\varphi}{\partial t}+\phi  =  0\,  . \label{ebaro}
\end{eqnarray}
The latter, Equation\,(\ref{ebaro}), is the classical Bernoulli
equation for compressible, unsteady flows without viscosity. Without loss of generality, we have 
incorporated a possible time-dependent integration constant into the function $\varphi$.
This equation
needs further integration, as the mass conservation Equation\,(\ref{mbaro}) must also
be fulfilled.
The study of possible temporal evolutions of the Euler Equation\,(\ref{epf}), 
together with the mass continuity equation Equation\,(\ref{mbaro}) is subject of 
Section~(\ref{septidepwograv}).

In analogy to the incompressibility representation ($\rho=\rho
(\psi)$), we introduce the affinity between isosurfaces of
the velocity potential and density, namely $\rho=\rho(\varphi)$, where $\rho'=\rho'(\varphi)=d\rho/d\varphi$.
Adding an explicit time dependence to the mass continuity Equation\,(\ref{mbaro}) results in the time-depending continuity equation
\begin{equation}
\frac{\partial\varphi}{\partial t}+\left({\boldsymbol\nabla}\varphi\right)^2 +
\frac{\rho}{\rho'}\,\Delta\varphi = 0\, .\label{sg0}
\end{equation}
Using this equation, we substitute either the quadratic derivative of the potential $\varphi$ or its time derivative in the Euler Equation\,(\ref{ebaro}) and obtain 
\begin{equation}
G(\varphi) +\frac{1}{2}\,\frac{\partial\varphi}{\partial t}
-\frac{\rho}{2\rho'}\,\Delta\varphi + \phi = 0
\label{sg1} 
\end{equation}
respectively
\begin{equation}
G(\varphi) - \frac{({\boldsymbol\nabla}\varphi)^2}{2}
 - \frac{\rho}{\rho'}\,\Delta\varphi + \phi = 0
 \label{sg}
\end{equation}
with 
\begin{eqnarray}
G(\rho(\varphi))=\int\frac{dp}{\rho(p)}=\int\frac{dp(\rho)}{d\rho}\,
\frac{d\rho}{d\varphi}\,\frac{d\varphi}{\rho(\varphi)}\, .\label{sg2}
\end{eqnarray}
%
The function $G$ can be identified as the specific enthalpy.

Equation\,(\ref{sg1}) is a non-linear diffusion equation, similar to the non-linear Schr\"odinger
equation, and Equation\,(\ref{sg}) is a convection-diffusion-type equation.

In the following we first discuss a special case with a
specific time dependency where only $\varphi$ is time dependent and $\rho$ is stationary and depends only on the spatial parts of $\varphi$ (Section~\ref{septidepwograv}), whereas in Section~\ref{cht} we return to the general case of a density
depending explicitly on the time-dependent velocity potential.

\subsection{Separable time dependence and neglection of gravity}
\label{septidepwograv}

We want to search for the general time dependence of (circumstellar) flows
in order to find out whether instabilities or 
collapse processes necessarily occur, even if the system stipulates a stationary density at far 
distances from the central star as observations suggest. In particular, we are interested
in an expansion of a solution around an equilibrium state, and this can be naturally achieved  
by the separation of the temporal and spatial parts of the corresponding fields. Therefore, we assume 
a separability of time dependency for the pressure and velocity potential, consider a 
stationary density, depending on the spatial part of the velocity potential, and
neglect gravity\footnote{The influence of gravity can be neglected for cases in which the gravitational force is considerably smaller than the pressure gradient, which is fulfilled at 
distances of circumstellar rings far from the star and pressure gradient lengthscales across the ring small compared to the distance of the ring.}, i.e.
$p=p_0(t)\, p_1(\varphi_1)$, $\varphi=\varphi_0(t)\,\varphi_1(x,y,z)$ and $\rho=\rho(\varphi_1)$.
The chosen time-separability for the pressure enables a barotropic law at any time. Inserting this separability ansatz into the conservation of mass equation for stationary density, Equation\,(\ref{mbaro}), we obtain the relation
\begin{equation}
({\boldsymbol\nabla}\varphi_{1})^{2} = -\frac{\rho(\varphi_1)}{\rho'(\varphi_1)} \Delta\varphi_1\, .
\label{eq40}
\end{equation}
With the 3D velocity potential $\textbf{\textit{v}}={\boldsymbol\nabla}\varphi$, and neglecting 
gravity, Equation\,(\ref{epf}) reduces to
\begin{equation}
\frac{\partial\textbf{\textit{v}}}{\partial t}+{\boldsymbol\nabla}\frac{\textbf{\textit{v}}^2}{2}
=-\frac{{\boldsymbol\nabla} p}{\rho}\, .
\end{equation}
Inserting the separation ansatz for the pressure and the velocity potential leads to
\begin{equation}
{\boldsymbol\nabla}(\dot{\varphi}_{0}(t)\varphi_{1}) + {\boldsymbol\nabla}\!
\left[\frac{\varphi_{0}^{2}(t)}{2} ({\boldsymbol\nabla}\varphi_{1})^{2}\right]
= -\frac{{\boldsymbol\nabla} \left(p_{0}(t) p_{1}(\varphi_{1})\right)}{\rho(\varphi_{1})}\, .\label{bla}
\end{equation}
The right-hand side can be written as
\begin{equation}
\frac{{\boldsymbol\nabla} \left(p_{0}(t) p_{1}(\varphi_{1})\right)}{\rho(\varphi_{1})} = {\boldsymbol\nabla}\left[p_{0}(t) \int\frac{p_{1}'(\varphi_{1})d\varphi_{1}}{\rho(\varphi_{1})}\right]\, .
\end{equation}
With this, Equation\,(\ref{bla}) can be formally integrated, and if we replace the term $({\boldsymbol\nabla}\varphi_{1})^{2}$ by Equation\,(\ref{eq40}), we obtain
\begin{equation}
p_{0}(t)\int \frac{p_{1}'(\varphi_{1})\,d\varphi_{1}}{\rho(\varphi_{1})}
-\frac{\varphi_{0}^{2}(t)}{2}\,\frac{\rho(\varphi_{1})}{\rho'(\varphi_{1})}\,\Delta\varphi_{1}+
\dot{\varphi}_{0}(t)\varphi_{1} =0\, . \label{Bern}
\end{equation}
We define
\begin{equation}
\tilde{F}:= \int \frac{p_{1}'(\varphi_{1})\,d\varphi_{1}}{\rho(\varphi_{1})}\, . \label{Ftilde}
\end{equation}
As one can see by inspection of Equation\,(\ref{sg2}), $\tilde{F}$ is up to a factor $p_0(t)$ identical to the specific 
enthalpy $G$. Inserting Equation\,(\ref{Ftilde}) into Equation\,(\ref{Bern}), we get
\begin{eqnarray}
p_{0}-\varphi_{0}^{2}\,\frac{\rho}{\rho'}\,
\frac{\Delta\varphi_{1}}{2\tilde{F}}+\frac{\dot{\varphi}_{0}\varphi_{1}}{\tilde{F}} &= &0\, \label{separab0} \\
\Leftrightarrow\qquad\frac{p_{0}}{\dot{\varphi}_{0}}-\frac{\varphi_{0}^{2}}{\dot{\varphi}_{0}}\,\frac{\rho}{\rho'}\,
\frac{\Delta\varphi_{1}}{2\tilde{F}}+\frac{\varphi_{1}}{\tilde{F}} &= &0\, . \label{separab1}
\end{eqnarray}
Equation\,(\ref{separab1}) is only valid for $\dot{\varphi}_{0} \neq 0$. To achieve the separability 
of this equation into spatial and time-dependent parts, we factorize the first terms in 
Equation\,(\ref{separab1}) and recognize that it is indispensable that the term 
$\frac{p_{0}}{\varphi_{0}^2}$ must be constant. We label this constant as $c_{1}$ and obtain 
\begin{equation}
\frac{\varphi_0^2}{\dot{\varphi_0}}\left(c_1-\frac{\rho}{\rho'}\,\frac{\Delta\varphi_{1}}{2\tilde{F}}\right)
+\frac{\varphi_1}{\tilde{F}} = 0\label{sep}\, .
\end{equation}
Equation\,(\ref{sep}) can only be solved if $\frac{\varphi_{0}^2}{\dot\varphi_0} = \varphi_{00} = 
\textrm{const}$. The integral of this differential equation is
\begin{equation}
\varphi_0 = \frac{\varphi_{00}}{t_0-t}\, , \label{eq49}
\end{equation}
delivering a finite-time singularity for the time-dependent part of the velocity potential.
The constant $\varphi_{00}$ describes the amplitude of the velocity potential imprinted onto the 
spatial part of the velocity potential for the time $t=0$.

For the spatial part we find 
\begin{equation}
 2\frac{\varphi_{1}}{\varphi_{00}}\frac{\rho'}{\rho}
+2\, c_1\tilde{F}\frac{\rho'}{\rho}=\Delta\varphi_1\, . \label{separab2}
\end{equation}

%
Equation\,(\ref{separab2}) is basically a nonlinear elliptic partial differential equation  whose solution reflects the 
spatial part of the scalar velocity potential $\varphi_{1}$ and thus determines the structure of the entire 
time-dependent flow. To solve this equation requires the knowledge of the specific enthalpy $\tilde{F}$ respectively 
$G$ which results from the specification of the density function $\rho(\varphi_{1})$ along with pressure dependency
$p_{1}(\rho)$ or explicitly $p_{1}(\varphi_{1})$. 

A detailed analysis of this equation is beyond the scope of the current paper, however, we wish to 
draw the attention to the fact that, if the time separability is requested along with a time 
independent density distribution (stationarity), then according to Equation\,(\ref{eq49}) for 
$\varphi_{00} > 0, t_0 > 0$ and $t_0 > t$ the system inevitably develops into a finite-time 
singularity for $t\rightarrow t_0$, whereas in the limit $t\rightarrow -\infty$ the velocity 
and pressure decay.

For a simplified case we can find a subspace. For example, for the case $\dot{\varphi}_{0} = 0$, 
Equation\,(\ref{separab0}) takes the following form
\begin{equation}
p_{0}-\varphi_{0}^{2}\frac{\rho}{\rho'}\,
\frac{\Delta\varphi_{1}}{2\tilde{F}} = 0\, . \label{separab01}
\end{equation}
This equation can only be fulfilled for $p_{0}=\textrm{const}$. In this case, the equation takes the
form of a quasi-linear partial differential equation
\begin{eqnarray}
\Delta\varphi_{1}=2\,\frac{\rho'(\varphi_1)}{\rho(\varphi_1)}\frac{p_0}{\varphi_0^2}\, \tilde{F}(\varphi_1)\, , \label{nlp}
\end{eqnarray}
which can also be derived directly from the non-linear Schr\"odinger Equation\,(\ref{sg1}).
Equation\,(\ref{nlp}) is a non-linear Poisson equation for which exact analytical solutions are known
in 2D.

\subsection{General time dependent approach and generalized Cole--Hopf transformation}
\label{cht}

We return now to the system of non-linear diffusion equations (Equations\,(\ref{sg0})--(\ref{sg2})) which we 
want to reformulate in a more compact form. For this, we use a generalized form of the Cole-Hopf transformation. 
The basic form of this transformation type has been invented by \cite{Hopf} and \citet{Cole}.

In the case of non-linear diffusion equations the spatial part of the differential operators can have the form
\begin{equation}
a\Delta\varphi+b({\boldsymbol\nabla}\varphi)^2=c \quad\land\quad a,b,c\,\,\textrm{functions}\, .
\label{cdi1}
\end{equation}
Let us assume $\varphi=F(\Lambda)$, we can rewrite Equation\,({\ref{cdi1}}) as
\begin{eqnarray}
a\left[F''(\Lambda)({\boldsymbol\nabla}\Lambda)^2+F'(\Lambda)\Delta\Lambda\right]
+b \left(F'(\Lambda)\right)^2 ({\boldsymbol\nabla}\Lambda)^2=c \, , \label{trafo}\nonumber\\
\end{eqnarray}
where the primes at $F$ denote derivatives with respect to $\Lambda$. The transformation
$\varphi=F(\Lambda)$ should be now done in such a way that terms with $({\boldsymbol\nabla}\Lambda)^2$ are eliminated. This demand can be formulated as
\begin{eqnarray}
a F''(\Lambda)+b \left(F'(\Lambda)\right)^2=0 \, .
\label{ch1}
\end{eqnarray}
To solve Equation\,(\ref{ch1}), it is necessary that the function $a/b$ only
explicitly depends on $\Lambda$. The transformation $F$ can be derived by
integration of the condition Equation\,(\ref{ch1}) which has the general solution 
\begin{eqnarray}
F=\displaystyle{\int \frac{d\Lambda}{\int \frac{b}{a}\, d\Lambda }}
\,\,\,\Leftrightarrow\,\,\, \Lambda=\int \exp{\left(\int \frac{b}{a}\, d\varphi\right)}
\, d\varphi \, .
\label{ch2}
\end{eqnarray}
With the condition (Equation\,(\ref{ch1})),  Equation\,(\ref{trafo}) reduces to
\begin{equation}
a F'(\Lambda)\Delta\Lambda = c\, . \label{trafo2}
\end{equation}
From a comparison of coefficients between Equation\,(\ref{sg0}) and Equation\,(\ref{cdi1}), we can 
derive the functions $a, b$ and $c$
\begin{equation}
a = \frac{\rho}{\rho'} \qquad \textrm{and} \qquad  b = 1 \qquad \textrm{and} \qquad c = 
-\frac{\partial\varphi}{\partial t}\, .
\end{equation}
Inserting the general solution (Equation\,(\ref{ch2})) along with the functions $a, b, c$ into Equation\,(\ref{trafo2}) leads to
\begin{eqnarray}
\varphi(\Lambda) &&=\int \left[\int
\frac{\rho\, d\Lambda}{\rho'}\right]^{-1} d\Lambda
\,\,\Leftrightarrow\,\, \Lambda(\varphi)=\int \rho\,  d\varphi
\nonumber\\
\nonumber\\
\Rightarrow\,\, \Delta\Lambda &&= -\frac{\partial\Lambda}{\partial t}\,\frac{\rho'}{\rho}\, .
\label{diff}
\end{eqnarray}
%
The last equation of Equation\,(\ref{diff}) is a nonlinear diffusion equation and is the Cole-Hopf 
transformed mass continuity equation.
For this kind of non-linear diffusion or Schr\"{o}dinger equations 
(Equation\,(\ref{diff}) and Equation\,(\ref{sg1})) it is known that they can have blow-up solutions
\citep[see, e.g.,][]{Galaktionov}.

Next, we transform Equation\,(\ref{sg}). Due to the mathematical similarity of 
Equation\,(\ref{sg}) and Equation\,(\ref{cdi1}) we can use the transformation Equation\,(\ref{ch2}) for
\begin{equation}
a = -\frac{\rho}{\rho'} \qquad \textrm{and} \qquad  b = -\frac{1}{2}\, .
\end{equation}
This leads to
\begin{eqnarray}
&& G\left(F\left(\Lambda\right)\right)-\frac{\rho}{\rho'}\, F'(\Lambda)\,\Delta\Lambda+\phi=0
\nonumber\\
\Rightarrow\quad && G\left(F\left(\Lambda\right)\right)-\frac{\rho}{\rho'}\,
\left[\int\,\frac{\rho'}{2\rho}\, d\Lambda\right]^{-1}\,\Delta\Lambda+\phi=0
\nonumber\\
\Rightarrow\quad && G\left(F\left(\Lambda\right)\right)-\frac{\rho}{\sqrt{\rho}\rho'}
\,\Delta\Lambda+\phi=0\, \nonumber\\
\Rightarrow\quad && G\left(F\left(\Lambda\right)\right)-\left[\frac{d{\rho}}{d\Lambda}\right]^{-1}
\Delta\Lambda+\phi=0\, .
\label{sg3}
\end{eqnarray}
The advantage of carrying out the transformation is that Equation\,(\ref{sg3}) does not contain
time derivatives and quadratic terms anymore, and that it is again a non-linear Poisson equation 
which can be solved with standard procedures. The only problematic and non-trivial issue is that 
$\rho(\varphi)$ must be chosen adequately in order to find solutions of the equation, based on which 
it will be possible to obtain the velocity field, pressure and density distribution.

\section{Solutions of Euler equation for special, persistent geometrical flow patterns}\label{sect:flow}

Now we drop the condition for incompressible or irrotational flows and treat the hydrodynamical 
problem of non-linear and non-local instabilities from a more general point of view.  
It is possible to implement the gravitation of the star into the pressure in case that either
$p$ can be considered as a function of $\rho$ (barotropic fluid), or $\rho$ as a function of $\phi$, 
although in the former case it will then be necessary to slightly modify the equation of motion. 
Without significant loss of generality, we therefore use only \textit{one} pressure gradient force and 
write the Euler equation for the stellar wind in the following form  

\begin{equation}
{\boldsymbol\nabla} p=-\rho\left(\textbf{\textit{v}}{\boldsymbol\cdot\boldsymbol\nabla}\right)\textbf{\textit{v}} 
-\rho\frac{\partial\textbf{\textit{v}}}{\partial t}\, .
\label{eu1}
\end{equation}
For this, the equilibrium solution, which one may also consider as the \lq ground state\rq~of the system, is
\begin{eqnarray}
{\boldsymbol\nabla} p_0=-\rho_0\left(\textbf{\textit{v}}_0{\boldsymbol\cdot\boldsymbol\nabla}\right)\textbf{\textit{v}}_0 \quad\land\quad
{\boldsymbol\nabla\boldsymbol\cdot}\left(\rho_{0}\textbf{\textit{v}}_{0}\right)=0\, .
\label{eu2}
\end{eqnarray}
If we assume the separability for the pressure $p=p_{1}(t)\, p_{0}(\textbf{\textit{x}})$ and  analogously also for the density 
$\rho = \rho_{1}(t)\rho_{0}(\textbf{\textit{x}})$ and for the velocity field $\textbf{\textit{v}}=v_{1}(t)\,\textbf{\textit{v}}_0 (\textbf{\textit{x}})$, 
then we will show in the following that for some choices of constraints there can exist regular solutions, and other choices can lead to a blow-up of
the stationary nonlinear spatial solution.

The separation implies that locally, i.e. inside some small subset of a
time interval, the fluid variables are not to be considered as fast
evolving, around a known stationary wind solution.
The mass continuity equation can then be written as 
\begin{equation}
\rho_{0}\dot{\rho}_{1}+\rho_{1} v_{1}\,{\boldsymbol\nabla\boldsymbol\cdot}\left(\rho_{0}\textbf{\textit{v}}_{0}\right)
=0\, , \label{mc2}
\end{equation}
and with the conditions of Equation\,(\ref{eu2}) it follows that
\begin{eqnarray}
\dot{\rho}_{1}=0 & \qquad \Rightarrow \qquad & \rho_{1}=1
\end{eqnarray}
without loss of generality.

The Euler equation (\ref{eu1}) together with the initial condition Equation (\ref{eu2}) can be expressed in the form given in Equation\,(\ref{eu2a}). Taking first
the divergence operator on both sides (${\boldsymbol\nabla\boldsymbol\cdot}$), and second the curl operator
(${\boldsymbol\nabla}{\boldsymbol\times}$), one receives
\begin{eqnarray}
&& p_1\,{\boldsymbol\nabla} p_0 = v_1 ^2\,{\boldsymbol\nabla} p_{0}-\rho_0 \dot{v}_{1} \textbf{\textit{v}}_{0}
\label{eu2a}\\
\Rightarrow\quad && \Delta p_0=0\quad\land\quad {\boldsymbol\nabla}{\boldsymbol\times}\left(\rho_0
\textbf{\textit{v}}_0\right)={\boldsymbol 0}\\
\Rightarrow\quad && \exists\,\varphi_{0},\, \Delta\varphi_{0}=0\, , v_{00}=\textrm{const.},
\,\textrm{with}\nonumber\\
&& 
 p_0=v_{00} \varphi_0\,\,\land\,\, v_{00}{\boldsymbol\nabla}\varphi_0\equiv
v_{00}\rho_{0}\textbf{\textit{v}}_0={\boldsymbol\nabla} p_{0} \label{blowup}\\
\Rightarrow\quad && p_1-v_1^2=-\frac{1}{v_{00}} \dot{v}_1 \, .\label{blowup1}
\end{eqnarray}
The condition $v_{00}<0$ is subject to the physical approach that the direction of the velocity should be anti-parallel to the pressure gradient in the relation
between the mass-current vector $\rho_0\,\textbf{\textit{v}}_0$, and the pressure gradient
Equation\,(\ref{blowup}). The assumption that $v_{00}<0$ reflects the decrease of the pressure from upwind to downwind direction, which usually drives a stellar wind or
emulates \lq decretion\rq.

Equation\,(\ref{blowup1}) represents an ordinary differential equation for the time
dependent dynamics of the flow, based on the equilibrium solution Equation\,(\ref{eu2}). 
Already under simplified assumptions, like $p_1=\textrm{const.}$, we can find solutions with 
finite-time singularities (blow-up solutions), but these solutions originate from a finite-time 
singularity and converge towards the equilibrium solution. In contrast to the decretion solution, the 
accretion solution would originate from the equilibrium solution and converge to a finite-time 
singularity. In the case $p_1=0$ it can clearly be 
seen and easily calculated that $v_1\propto 1/(t_0-t)$. This solution must be excluded, as this would 
imply that the time perturbation reduces the complete pressure to zero for all times, implying 
a \lq pressure-less fluid\rq, i.e., a completely force-free flow.

For $p_1 = \textrm{const.} \neq 0$ we calculate the formal solution:
\begin{equation}
t-t_0 =\int\limits_{v_{10}}^{v_{1}}\frac{-\frac{1}{v_{00}}\, d \tilde{v}_1}{p_1 - \tilde{v}_1^2}
=\frac{1}{2 |v_{00}|\sqrt{p_1}} \ln \frac{\tilde{v}_1 + \sqrt{p_{1}}}{\tilde{v}_1 - \sqrt{p_{1}}} \Bigg|_{v_{10}}^{v_{1}} 
\end{equation}
leading to
\begin{equation}
|v_{00}|\sqrt{p_1} (t-t_0) = \frac{1}{2} \left(\ln \frac{v_{1}+\sqrt{p_{1}}}{v_1 - \sqrt{p_{1}}}-\ln \frac{v_{10}+\sqrt{p_{1}}}{v_{10} - \sqrt{p_{1}}} \right)\, .
\label{zeit1}
\end{equation}
%
For $t>t_0$ and $v_1, v_{10} > \sqrt{p_1}$ Equation\,(\ref{zeit1}) can be reformulated as
\begin{eqnarray}
\!\!\!\!\!\!&& \textrm{ar}\!\coth{\left(\frac{v_1}{\sqrt{p_1}}\right)} = |v_{00}|\sqrt{p_1} (t-t_0) + C^{2} \\
\!\!\!\!\!\!\Rightarrow\quad  & & v_1(t)  =  \sqrt{p_1} \coth{\left(|v_{00}|\sqrt{p_1} (t-t_0) + C^{2} \right)} 
\label{zeit2}
\end{eqnarray}
with 
\begin{equation}
C^{2} = \textrm{ar}\!\coth{\left(\frac{v_{10}}{\sqrt{p_1}}\right)}\, .
\end{equation}
Then, the solution of Equation\,(\ref{zeit2}) implies that the velocity decays with time and relaxes 
into an equilibrium state.

If we assume that $p_1$ is not constant anymore but $p_1=p_1(t)$ can be imposed, then 
Equation\,(\ref{blowup1}) turns into a ordinary differential equation of Riccati 
type. 

As the problem is nonlinear anyway, the function $p_1$ can be regarded, especially in this 
situation of a dependency of only one variable, namely $t$, as function of $v_1$, i.e.
$p_1=p_1(v_1)$. For certain constraints we are able to show that equilibria could converge to
blow-up solutions. The assumption that $p_1(v_1) \geq 0$ guarantees that the pressure is always 
positive. For a continuous outflow we must also guarantee that $\dot{v}_{1}$ is always larger than 
zero, if we exclude oscillatory phases of acceleration and deceleration. These conditions are 
fulfilled by $p_1-v_1^2 \geq 0$ (see Equation\,(\ref{blowup1})). Thus, it is further necessary that $p_1$ is not only positive,
and a sufficient criterion to guarantee the validity of $p_1-v_1^2 \geq 0$ is that there exists a 
$p_{10} > 1$ and $p_1=p_{10} v_{1}^2$ such that 
\begin{equation}
p_{10} v_{1}^2 - v_{1}^2 =  \frac{1}{|v_{00}|} \dot{v}_{1}\, .
\end{equation}
Integration of this equation results in
\begin{equation}
v_1(t) = \frac{1}{(p_{10}-1)|v_{00}|(t_0-t) + \frac{1}{v_{10}}}
\end{equation}
with $v_1(t_0) = v_{10}$. 
For $t \rightarrow t_{\rm crit}$ the system develops a blow-up, where 
\begin{equation}
t_{\rm crit} = \frac{1}{v_{10}(p_{10}-1)|v_{00}|} + t_0\, .
\end{equation}
To summarize, the choice of $p_{1}(v_1)$ determines the temporal course of the flow and we have shown
two contrasting examples, namely a decaying perturbation (for $p_{1} = \textrm{const.}$) and a 
perturbation that causes a blow-up of the system (for $p_{1} = p_{10} v_{1}^{2}$). More complex 
choices of $p_{1}(v_1)$, e.g. with higher orders of the polynomial, can lead 
to correspondingly diverse results such as multiple blow-ups and decays. Other time courses are quite 
possible and will occur, but one should keep in mind that the blow-up solutions can be damped by 
normal and anomalous dissipative processes, leading to a regular time-dependent behavior. However, 
the blow-up solutions indicate that non-linear 
instabilities can develop in quasi-ideal systems and hence can lead to abrupt changes of the typical 
temporal scales.

\section{Discussion and conclusions}\label{sect:concl}

The environments of certain types of evolved massive stars, such as the B[e] supergiants, 
display indications for disks, or multiple ring-like structures of yet unknown origin, which 
cannot be reconciled with the classical theory of a (viscous) outflowing disk 
\citep[e.g.,][]{1991MNRAS.250..432L, 2001PASJ...53..119O, 2018A&A...613A..75K}. These 
configurations can be either steady \citep{2016A&A...593A.112K, 2023Galax..11...76K} or display 
temporal variability \citep{2018MNRAS.480..320M, 2018A&A...612A.113T}. 
Therefore we have two problems, namely (i) either long time scales (equilibria), or systems close to 
or converging to equilibrium states, or (ii) we need short time scales to explain sudden appearance of 
new structures \citep[such as detected by][]{2012MNRAS.426L..56O} resulting from non-linear 
instabilities (occurring either within the circumstellar matter or already in the stellar 
atmosphere) which might be connected to collapse processes (finite-time singularities). To study
the high diversity of geometrical structures and their possible formation mechanism 
we recourse to elementary mathematical tools of classical hydrodynamics such as linear and non-linear 
potential theory. 

For the first case of long time scales, complex geometries of streamlines are known and can be 
constructed from potential representation of equilibrium fields, e.g. potential fields (Laplace 
equation), or by non-linear solutions of Grad-Shafranov-type equations, where closed and open 
streamline configurations (arcs, rings, radial structures) are possible 
\citep[see, e.g.,][]{2013A&A...556A..61N, 2014ASTRP...1...51N}. In our analysis 
we include flows and gravity and find that the equations are of similar structure, which means that 
they can be solved in an analogous manner. We could prove that for a non-pressureless gas we can find
quasi-Keplerian or even Keplerian rotation of the circumstellar matter.

To bridge the gap from these equilibrium structures to short time scales a \lq 
homotopic\rq~modulation, i.e. a dilatation of the quasi-equilibria is performed, leading eventually  
to restrictions for the equilibrium values and to 
ordinary differential equations as constraints for the time dependencies of the fluid equations.
These ordinary differential equations in time lead either to regular (decay or growth, or relaxation)
or non-regular time dependencies where the latter includes finite-time singularities, 
which are known from (magneto-)hydrodynamical systems \citep[e.g.,][]{1996PhPl....3.4281K, 
2008ASTRA...4....7N}, and a broad overview about mathematical solution techniques for different 
physical problems can be found in \citet[][and references therein]{galaktionov2006exact}. 

For future investigations, non-separable time dependencies can be taken into account to get a better 
understanding of the variability and changes of the circumstellar matter.  Moreover, for studies of 
the conditions under which rings and arcs can form within the circumstellar 
material it will be important to analyze in more detail the influence of different conformal mappings
or multipole components, being able to create arcs, rings, and spiral-arm like structures. Such an 
analysis will then allow to derive the properties of the gas (density, temperature, and emissivity) 
which can be compared to observed quantities.

\medskip

We wish to thank Prof.~Dr.~Wolfgang Glatzel for the warm hospitality during our visit at the Institut 
f\"{u}r Astrophysik (IAG), Georg-August-Universit\"{a}t G\"{o}ttingen, and for the many fruitful 
discussions.
This research made use of the NASA Astrophysics Data System (ADS).
We acknowledge financial support from the Czech Science Foundation (GA\,\v{C}R, grant 
number 20-00150S). The Astronomical Institute Ond\v{r}ejov is supported by the project RVO:67985815. 
This project has received funding from the European Union's Framework Programme for Research and 
Innovation Horizon 2020 (2014-2020) under the Marie Sk\l{}odowska-Curie Grant Agreement No. 823734.

\bibliography{ms}{}
\bibliographystyle{aasjournal}

\end{document}